\documentclass[12pt,twoside,espcrc1]{article}
\usepackage{epsfig}
\usepackage{espcrc1}
\psfigdriver{dvips} 
\begin{document}
\pagestyle{empty}
\oddsidemargin 0.0cm
\evensidemargin 0.0cm
\normalsize

\vspace*{1.5cm} \noindent {\large Anisotropic Azimuthal Distributions of 
Identified Particles in Au+Au \\ Collisions at 11.5 A GeV/c} \\ \ 

\noindent { Jean Barrette$^a$ for the E877 Collaboration\footnote{
BNL, GSI, Universit\"at Heidelberg, Idaho Nat. Eng. Lab., 
McGill University, 
University of Pittsburgh, SUNY at Stony Brook, 
University of S\~ao Paolo, Wayne State University
} \\

\noindent $^a$Department of Physics, McGill Univ., Montreal, Canada H3A 2B2 \\

\vskip 0.1cm


Anisotropy in the azimuthal distribution of the 
reaction products in heavy-ion collisions is an important 
observable which, in particular, contains information on the 
evolution and amplitude of the pressure gradient in 
the compressed nuclear matter formed in the collision 
and thus on its equation of state.  It was first observed in 
ultra-relativistic collisions in 1994 \cite{FlowPRL} and
is now considered one of the best 
tools to elucidate the dynamics of heavy-ion collisions. 

In this contribution we present recent results 
on the study of azimuthal distributions of produced particles
in Au+Au 
collisions at 11.5A GeV/c obtained from the analysis
of the last E877 data taking run in 1995. 
Other E877 results on the flow of particles,
particularly on the elliptical flow,
are also presented in the contribution 
of K. Filimonov at this conference \cite{qm99kirill}. \\

\noindent {\bf 1. EXPERIMENTAL METHOD} \\

The E877 set-up is discussed in detail in 
\cite{DeeCentral,etmult,protplane,lightplane}. 
The azimuthal anisotropy of 
the particle production is studied by 
means of Fourier analysis of the azimuthal 
particle distributions measured with respect to the reaction plane orientation
\cite{Fourier} . 
In this representation, 
the dipole (v$_1$) and the quadrupole (v$_2$) coefficients in the 
Fourier decomposition represent the shift (directed flow) 
and the eccentricity (elliptical flow) of the particle azimuthal 
distribution. 

In E877, the reaction plane angle is determined 
for every event using the direction of the 
transverse energy flow measured in 
calorimeters surrounding the target. 
The details of this procedure are given in \cite{etmult}.  
The measured anisotropies are 
corrected for the reaction plane resolution. \\

\noindent {\bf 2. RESULTS} \\

Results on the directed flow of protons, 
pions and light nuclei have been published 
previously \cite{protplane,lightplane}. 
It was shown, in particular, that in semicentral Au+Au collisions, protons 
exhibit strong directed flow that  
is well reproduced by RQMD if the effects of mean-field are included. 
Here we present results on flow measurements of other produced particles.


\begin{figure}
\epsfig{file=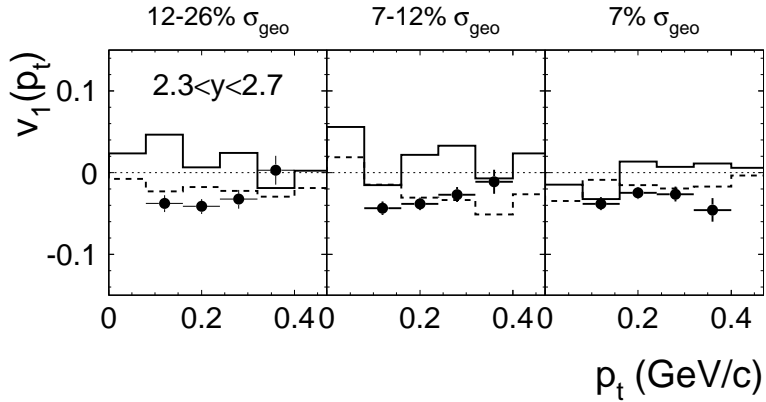,height=5.5cm,clip=,}
\begin{minipage}[c]{4cm}
{\vspace*{-7cm}
\hspace*{0.5cm}
\caption{\small $v{_1}(p{_t})$ of $K{^+}$ (solid circles). The histograms 
are the predictions of RQMD(2.3) run in cascade (dashed lines) and
mean field (full lines) modes.}}
\end{minipage}
\end{figure}

Model calculations show that the kaon flow 
is sensitive to the value of the kaon
potential in the nuclear medium \cite{LiPRL}. 
In Fig.~1 we present the measured directed
flow amplitude v$_1$ as a function of $p{_t}$ for K$^+$. 
The K$^+$ flow signal is weak but clearly negative, i.e
in the direction opposite to that of protons. 
The measured K$^+$ directed flow is in good agreement 
with the predictions of RQMD 2.3 run in  
cascade mode whereas 
the agreement is not as good when the effect of mean field 
is included.

In the same rapidity window
K$^-$ show, within error, a smaller flow 
than K$^+$ (See Fig.~4 in \cite{qm99kirill}).
This result is contrary to the predictions of RQMD 
which predict a larger negative flow for K$^-$ than for K$^+$.


It was shown that the $\Lambda$ flow in heavy-ion collisions is
sensitive to the  properties of $\Lambda$-hyperons
in dense matter \cite{LiLambda}.
Using information from new upstream tracking chambers added to the E877
set-up for the 1995 running period, we could 
isolate secondary decay vertices and identify 
$\Lambda$'s through the reconstruction of $p\pi{^-}$ pairs (Fig~2). 
Due to the limited angular 
acceptance of the E877 spectrometer 
we are mainly sensitive to $\Lambda$'s emitted at 
forward rapidity (y$>$2.2) and the total geometrical 
acceptance for the produced $\Lambda$'s is of the order of 10$^{-3}$. 
A further 
reduction of about a factor of ten in the efficiency for detecting 
$\Lambda$'s results from 
additional cuts
that are introduced in the analysis to reduce the background from
uncorrelated tracks.

\begin{figure} [b]
\hspace*{1.0cm}
\epsfig{file=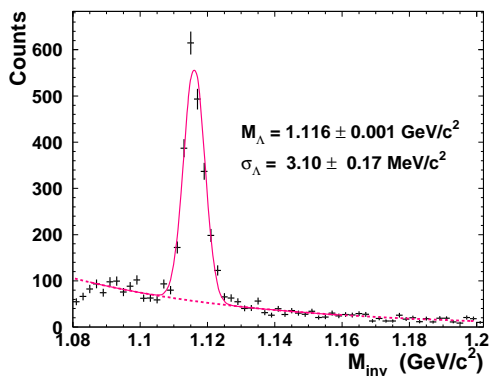,height=5.0cm,clip=,}
\begin{minipage}[c]{6cm}
{\vspace*{-7cm}
\hspace*{1.5cm}
\caption{\small Invariant mass spectrum of $p\pi{^-}$ pairs.}}
\end{minipage}
\end{figure}

The azimuthal distribution of $\Lambda$'s 
for semicentral collisions (Fig.~3) shows a clear positive 
anisotropy for the most forward rapidity window.
Interpretation of the v$_1$($p{_t}$)-dependence is limited due to low
statistics.
However,
one can conclude that the 
measured $p{_t}$ dependence is 
consistent with the predictions of 
RQMD(2.3) with, as in the case of K$^+$, a better 
description being observed for the pure cascade calculation.

\begin{figure} 
\hspace*{3.0cm}
\epsfig{file=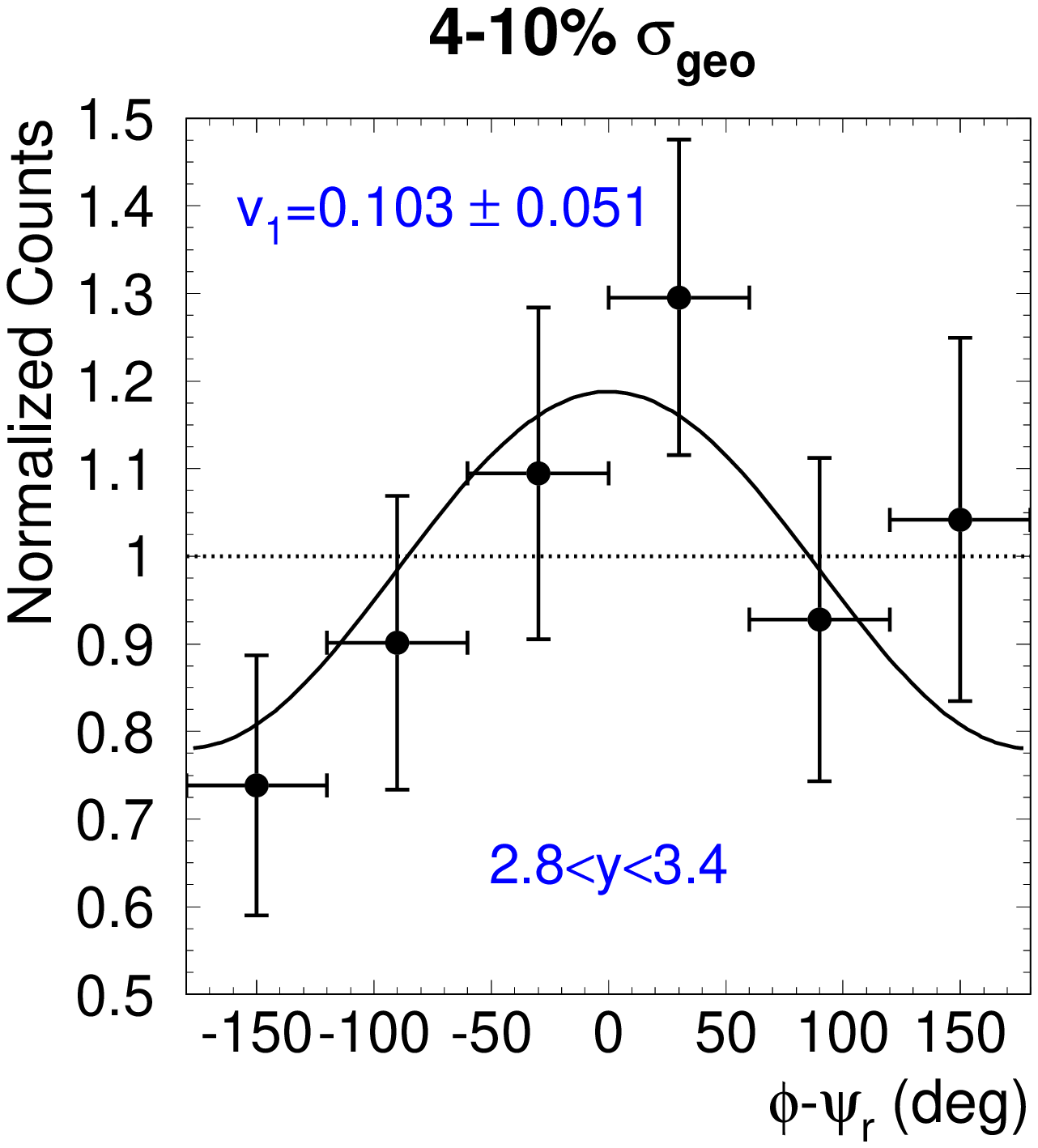,height=5.0cm,clip=,}
\hspace*{0.3cm}
\epsfig{file=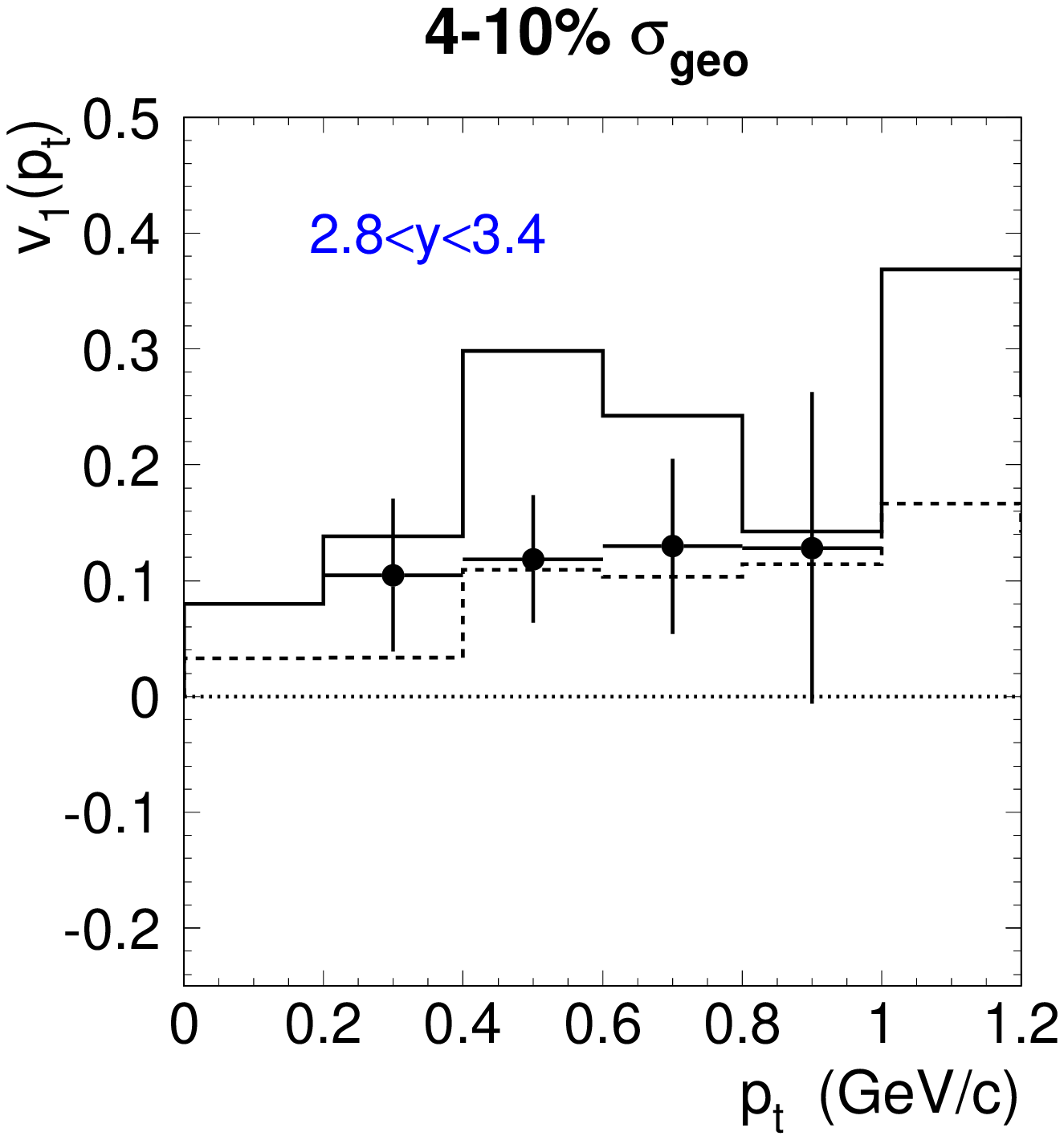,height=5.0cm,clip=,}
\vspace*{-0.5cm}
\caption{\small 
(left) Lambda azimuthal distribution at forward rapidity.
(right) Comparison of v$_1$($p{_t}$) of Lambdas 
with the predictions from RQMD(v2.3) run in cascade (dashed lines) 
and mean-field (full lines) modes.}
\end{figure}

%
%

The production of antiprotons at AGS 
energies is near threshold in the nucleon-nucleon rest frame. 
The yield and distribution of $\bar{p}$ 
is the result of both the production and subsequent 
annihilation. 
Antiprotons co-moving with the nucleons have 
a greater probability of annihilation and rescattering 
resulting in an anticorrelation with 
the nucleon directed flow leading to the so-called 
anti-flow of antiprotons in nuclear collisions predicted in \cite{Jahns}.

\begin{figure} [b]
\hskip 0.3 cm
\epsfig{file=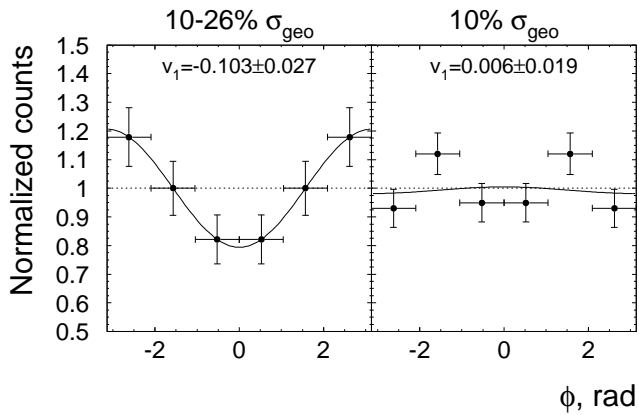,height=5.5cm,clip=,}
\begin{minipage}[c]{6cm}
{\vspace*{-7cm}
\hspace*{1 cm}
\caption{\small Antiproton azimuthal distribution at forward rapidity.}}
\end{minipage}
\end{figure}

The addition of the new tracking detectors has allowed 
a clean identification of $\bar{p}$ with however 
limited statistics due to the small $\bar{p}$ production cross section. 
Because of the amplitude of the spectrometer magnetic field selected in 1995  
the acceptance for $\bar{p}$ is close to 
mid rapidity where the directed flow signal is minimal. 
To optimize a possible flow signal we are presenting 
the measured azimuthal distribution of $\bar{p}$ emitted at rapidity 
y$>$1.8 which leaves about 400 $\bar{p}$ in the sample. 
The azimuthal distribution of antiprotons 
with respect to the reaction plane measured 
in the forward rapidity (1.8$<$y$<$ 2.2) 
for two centralities is shown in Fig~4. 
For clarity, 
taking into account that for reason of 
symmetry the distribution has to be symmetric 
about $\phi$=0, the data in the angular range [-$\pi$,0] have 
been averaged with the data in the range [0,+$\pi$] and 
the resulting data have been reflected about 
$\phi$=0. A pronounced minimum is observed at $\phi$=0 
for semicentral collisions
indicating that the antiprotons 
are, as predicted, strongly 
anticorrelated with the nucleons. 
The corresponding transverse momentum dependence of 
v$_1$ is presented in \cite{qm99kirill}. 
The observation of the large $\bar{p}$ antiflow confirms
that strong annihilation processes are 
involved in the dense nuclear matter and more 
precise data could provide a better understanding 
on how this annihilation is modified in the nuclear medium.

\begin{figure}
\hspace*{0.5cm}
\epsfig{file=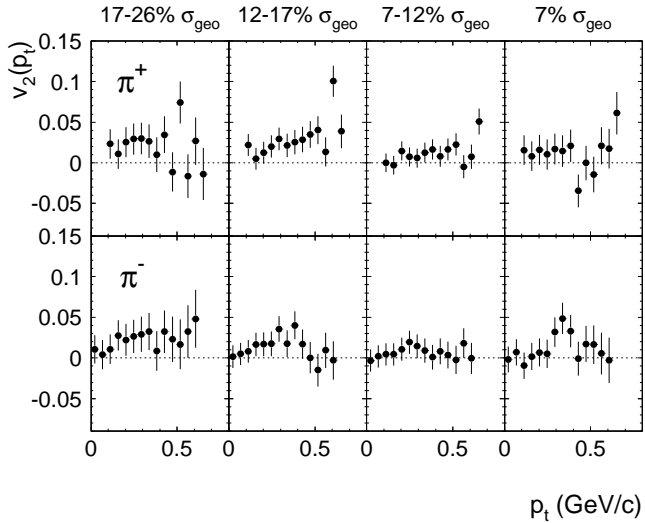,height=7.0cm,clip=,}
\begin{minipage}[c]{4cm}
{\vspace*{-9.0cm}
\hspace*{-1.0cm}
\caption{\small $v{_2}(p{_t})$ of $\pi{^+}$ 
and $\pi{^-}$  
in the rapidity range 2.8$<$y$<$3.2 
for different centrality bins.}}
\hspace*{-1.0cm}
\end{minipage}
\end{figure}

The larger statistics obtained in 1995 has allowed to obtain more extensive
data on the elliptic flow of produced particles.
Fig.~5 presents the measured $v{_2}(p{_t})$
of forward rapidity pions as function of centrality.
Both pion types exhibit a similar behaviour within experimental uncertainties.
Weak positive values of $v{_2}$ are observed with larger signals
for peripheral collisions. This corresponds to an elliptically shaped 
distribution with the major axis lying in the reaction plane.
This is opposite to what is observed at lower beam energies \cite{Brill}.

Financial support from US DoE, the NSF, the Canadian NSERC, 
is acknowledged.


\end{document}